\documentclass[preprint,12pt]{elsarticle}

\usepackage[ruled,vlined,linesnumbered,commentsnumbered]{algorithm2e}
\usepackage{amsmath,amssymb,amsfonts}

\def\FF {{\cal F}}
\def\II    {{\cal I}}
\def\TT {{\cal T}}

\def\si     {\sigma}
\def\eps {\epsilon}

\def\empt {\emptyset}
\def\sem  {\setminus}
\def\subs {\subseteq}

\def\f   {\frac}

\def\TCA       {{\sc $1$-CA}}
\def\CFC      {{\sc CFC}}
\def\SSCDS {{\sc SS-CDS}}
\def\TCDS {{\sc $2$-CDS}}

\newtheorem{theorem}{Theorem} 
\newtheorem{lemma}[theorem]{Lemma}
\newtheorem{corollary}[theorem]{Corollary}

\newenvironment{proof}{\noindent{\bf Proof:\/}}{\hfill $\Box$\vskip 0.1in}

\journal{Discrete Applied Mathematics}

\begin{document}

\begin{frontmatter}

\title{2-node-connectivity network design\footnote{Preliminary version appeared in \cite{N-W}.}} 

\author{Zeev Nutov}   \ead{nutov@openu.ac.il}        \address{The Open University of Israel}


\newcommand {\ignore} [1] {}


\begin{abstract}
We consider $2$-connectivity network design problems in which we are given a graph 
and seek a min-size $2$-connected subgraph that satisfies a prescribed property.
\begin{itemize}
\item
In the {\sc $1$-Connectivity Augmentation} problem the goal is to augment a connected graph 
by a minimum size edge subset of a specified edge set such that the augmented graph is $2$-connected.
We breach the natural approximation ratio $2$ for this problem,
and also for the more general {\sc Crossing Family Cover} problem.
\item 
In the {\sc $2$-Connected Dominating Set} problem, we seek a minimum size $2$-connected subgraph 
that dominates all nodes. We give the first non-trivial approximation algorithm 
with expected ratio $O(\sigma \log^3 n)$, where $\sigma=O(\log n \cdot\log\log n\cdot(\log\log\log n)^{3})$.
\end{itemize}

The unifying technique of both results is a reduction to the {\sc Subset Steiner Connected Dominating Set} problem.
Such a reduction was known for edge-connectivity, and we extend it to $2$-node connectivity problems. 
We show that the same method can be used to obtain easily polylogarithmic 
approximation ratios that are not too far from the best known ones for several other problems.
\end{abstract}

\begin{keyword}
$2$-connectivity, dominating set, approximation algorithm, symmetric crossing family
\end{keyword}

\end{frontmatter}

\section{Introduction} \label{s:intro}

We consider problems when we are given a graph, possibly with edge costs,
and seek a minimum size/cost $2$-connected subgraph that satisfies a given property.
Our first problem is the node-connectivity variant of the extensively studied {\sc Tree Augmentation} problem,
c.f. \cite{FJa,KT,K-sur,CJR,N,CKKK,EFKN-TALG,CN,KN-TAP,N-T,FGKS,GKZ,A,CTZ,TZ2,TZ}. 

\begin{center} \fbox{\begin{minipage}{0.98\textwidth}
\underline{\sc $1$-Connectivity Augmentation} ({\TCA}) \\
{\em Input:} \ \  A  connected graph $G=(V,E_G)$ and an edge set $E$ on $V$.  \\
{\em Output:} A minimum size edge subset $J \subs E$ such that $G \cup J$ is $2$-connected. 
\end{minipage}} \end{center}

The problem is NP-hard and admits approximation ratio $2$ \cite{FJa};
prior to our work no better ratio was known. 
In fact, we will consider a generic problem that, as we will show, includes {\TCA}. 
We need some definitions to present this problem. 
Two sets $A,B$ on a groundset $V$ {\bf cross} if $A \cap B \neq \empt$ and $A \cup B \neq V$. 
A set family $\FF$ is a {\bf crossing family} if $A \cap B,A \cup B \in \FF$ whenever $A,B \in \FF$ cross;
$\FF$ is a {\bf symmetric family} if $V \sem A \in \FF$ for all $A \in \FF$.
For example, the minimum edge-cuts of a graph $G$ form a symmetric crossing family, 
with the additional property that whenever $A,B \in \FF$ cross and $A \sem B,B \sem A$ are both non-empty, 
the set $(A \sem B) \cup (B \sem A)$ is not in $\FF$.
Dinitz, Karzanov, and Lomonosov \cite{DKL} showed that such a crossing family can be represented by $2$-edge cuts of a {\bf cactus} 
- a connected graph in which every block is a 
cycle.\footnote{When the edge-connectivity of $G$ is odd, the minimum edge-cuts not containing a specified node $r$ 
form a laminar family, and thus can be represented by a tree.}
Dinitz and Nutov \cite[Theorem 4.2]{DN} (see also \cite[Theorem 2.7]{N-Th})
extended this by showing that any symmetric crossing family can be represented 
by $2$-edge-cuts and specified $1$-node-cuts of a 
cactus.\footnote{An identical representation was announced later by Fleiner and Jord\'{a}n \cite{FJ}.}

An {\bf edge $f$ covers a set $A$} if $f$ has exactly one end in $A$, 
and an edge set $J$ covers a set family $\FF$ if every $A \in \FF$ is covered by some $f \in J$. 

\begin{center} \fbox{\begin{minipage}{0.98\textwidth}
\underline{\sc Crossing Family Cover} ({\CFC})  \\
{\em Input:} \ \ A symmetric crossing family $\FF$ on $V$ and an edge set $E$ on $V$. \\
{\em Output:} A  minimum size edge set $J \subs E$ that covers $\FF$.
\end{minipage}} \end{center}
The size of $\FF$ may be exponential in $|V|$, 
e.g., all proper subsets of $V$ form a crossing family. 
So here $\FF$ may not be given explicitly, and we just require 
that certain queries related to $\FF$ can be answered in time polynomial in $|V|$. 
A standard assumption is that for any $(s,t) \in V \times V$ and any edge set $J$,
we can find in polynomial time an inclusion minimal set $C^{st}  \in \FF$ with $s \in A$ and $t \notin A$
that is not covered by $J$, if such $C^{st}$ exists.
Another possibility is that we are given the \cite{DN} representation of $\FF$ that has size $O(|V|)$.

By \cite{DKL} (see also \cite{DN,FJ}), {\CFC} includes the {\sc Edge-Connectivity Augmentation} problem,
that seeks to increase the edge-connectivity of a given graph $G$ by adding to $G$ a minimum size edge set.
Here we observe that it also includes {\TCA}, and in particular prove the following.

\begin{theorem} \label{t:1}
If {\CFC} admits approximation ratio $\alpha$ then so is {\TCA}.
\end{theorem}

Next, we will reduce {\CFC} to the following problem. 

\begin{center} \fbox{\begin{minipage}{0.98\textwidth}
\underline{\sc Subset Steiner Connected Dominating Set} ({\SSCDS}) \\
{\em Input:} \ \  A  graph $H=(U,I)$ and an independent set $R \subs U$ of terminals.  \\
{\em Output:} A minimum size connected $R$-dominating set $J \subs U \sem R$ , namely, 
$H[J]$ should be connected and every $r \in R$ should have a neighbor in $J$.
\end{minipage}} \end{center}

It can be deduced from \cite{BN} (see also Section~\ref{s:3}) 
that the approximability of {\SSCDS} is the same as that of the {\sc Group Steiner Tree} problem,
that admits ratio $O(\log^3 n)$ \cite{GKR}.
However, {\sc SS-CDS} instances arising from the {\CFC} reduction 
have two properties, that make the problem much easier, 
by relating it to special instances of the {\sc Steiner Tree} problem,~see \cite{BGA}. 

\medskip

\noindent
\hspace*{0.3cm} {\bf Property 1:} The neighbors of every terminal $r \in R$ induce a clique. \\
\hspace*{0.3cm} {\bf Property 2:} Every non-terminal $u \in U \sem R$ has at most $2$ neighbors in $R$.

\begin{theorem} \label{t:2}
If {\SSCDS} with properties $1,2$ admits ratio $\alpha$, then so is {\CFC}.
\end{theorem}

When the preliminary version \cite{N-W} of this paper appeared, the best known ratio for 
{\sc SS-CDS} with properties $1,2$ was (roughly) $1.91$ due to Byrka, Grandoni, and Ameli \cite{BGA}.
Since then, the ratio was improved by Angelidakis, Denesik, and Sanit\'{a} \cite{ADS} to $1.892$.
Thus we get the following generic result, that was stated in \cite{ADS} for the particular case of {\TCA}.

\begin{corollary}
{\CFC} (and thus also {\TCA}) admits approximation ratio $1.892$.
\end{corollary}

Our main second problem is the $2$-connectivity variant of the {\sc Connected Dominating Set} problem,
c.f. \cite{GK99}, that is closely related to the {\sc Node-Weighted Steiner Tree} problem \cite{KR}.

\begin{center} \fbox{\begin{minipage}{0.97\textwidth}
\underline{\sc $2$-Connected Dominating Set} ({\sc $2$-CDS}) \\
{\em Input:} \hspace*{0.18cm} A  graph $G=(V,E)$. \\
{\em Output:} 
A $2$-connected subgraph $(S,F)$ of $G$ that dominates $V$ (namely, $S$ dominates $V$) and has minimum number of edges/nodes.
\end{minipage}} \end{center}

Note that since $|S| \leq |F| \leq 2|S|$ holds for any edge-minimal $2$-connected graph $(S,F)$ \cite{Mad-und},
the approximability of the objectives of minimizing the number of nodes and the number of edges are equivalent,
up to a factor of $2$.

{\TCDS} is $\Omega( \log n)$-hard to approximate \cite{GK98}, 
but prior to our work no nontrivial approximation ratio was known for it.
We need some definitions to present our results. 
Let $d_H(u,v)$ denote the distance between nodes $u$ and $v$ in a graph $H$.
Given a distribution $\{p_T:T \in {\cal T}\}$ over a subset ${\cal T}$ of spanning trees of a graph $G=(V,E)$, 
the {\bf stretch} of the distribution is ${\displaystyle \max_{u,v \in V}}  \sum_{T \in {\cal T}} p_T \frac{d_T(u,v)}{d_G(u,v)}$.
Let $\sigma=\sigma(n)$ be the lowest known stretch that can be achieved by a polynomial time construction of such 
a distribution for any graph on $n$ nodes. By \cite{ABN}, $\si=O(\log n \cdot\log\log n\cdot(\log\log\log n)^{3})$.

\begin{theorem} \label{t:3}
If {\SSCDS} admits ratio $\alpha$ then {\sc $2$-CDS} admits a polynomial time algorithm 
with expected approximation ratio $O(\alpha \cdot \si)$. 
\end{theorem}

Since {\SSCDS} admits ratio $O(\log^3 n)$ (see Section~\ref{s:3}), we get: 

\begin{corollary} \label{c:3}
{\TCDS} admits expected approximation ratio $O(\si \log^3 n)$.
\end{corollary}

Our main purpose is to make a progress on approximability of two old $2$-connectivity fundamental problems 
{\TCA} and {\TCDS}, c.f. \cite{K-sur} and \cite{DW}, respectively. 
But we also show that a similar method enables to obtain easily 
approximation ratios that are not too far from the best known ones for some problems
considered by Chekuri and Korula~\cite{CK}.
\begin{itemize}
\item
{\sc $2$-Connected $k$-Subgraph}: Given a graph $G$ with edge costs and an integer $k$,
find a min-cost $2$-connected subgraph of $G$ with at least $k$ nodes. 
We will show that this problem admits expected approximation ratio $O(\si \log k)$, 
almost matching the ratio $O(\log n\log k)$ of \cite{CK}.
\item
{\sc $2$-Connected Quota Subgraph}:
Given a graph $G$ with edge costs, node profits, and a quota $Q$,
find a min-cost $2$-connected subgraph of $G$ that has profit at least $Q$.
We will show that this problem 
admits expected ratio $O(\si \log n)$, almost matching the ratio $O(\log^2 n)$ of \cite{CK}.
\end{itemize}

\begin{table} 
\begin{center}
\hspace*{-0.65cm}
\tabcolsep=0.05cm
\begin{tabular}{|c|c|c|c|c|c|c|c|c|}  \hline   
  & {\bf connected}
  & \multicolumn{2}{c|}{{\bf $2$-edge-connected}}   
  & \multicolumn{2}{c|}{{\bf $2$-connected}}
\\\hline
{\bf problem}         
  &  
  & \ previous \ & \ this paper \ 
  & \ previous \ & \ this paper \ 
\\\hline 
{\sc \small $2$-CA}        & --                                       & $1.458$ \cite{GKZ}        & --                       & $2$ \cite{FJa}                     & $1.91$                 \\\hline 
{\sc \small $2$-CDS}          & $O(\log n)$ \cite{GK98}  & $O(\si \log n)$ \cite{BN} & --                       & --                                           & $O(\si \log^3 n)$ \\\hline \hline
{\sc \small $k$-Subgraph} & $O(\log k)$ \cite{BHL}     & $O(\log k)$ \cite{KS}      & --                       & $O(\log n\log k)$ \cite{CK} & $O(\si \log k)$     \\\hline
{\sc \small Quota}                & $O(\log n)$ \cite{BHL}    & $O(\log^2 n)$ \cite{CK}  & $O(\si \log n)$ & $O(\log^2 n)$ \cite{CK}       & $O(\si \log n)$	   \\\hline
\end{tabular}
\caption{Previous and our ratios for some $2$-connectivity network design problems.
Here $\si=O(\log n \cdot\log\log n\cdot(\log\log\log n)^{3}) \approx O(\log^3 n)$.
The ratios for plain connectivity are for the corresponding node weighted problems.}
\end{center}
\label{tbl:results} 
\vspace*{-0.38cm}
\end{table}

Our ratios and previous best known ratios are summarized in Table~1.
The first two rows address our main results.
The purpose of the other two rows is just to demonstrate the method,
that enables to obtain easily polylogarithmic approximation ratios that are not too far from the best known ones.

A unifying technique that we use in the proofs of Theorems \ref{t:2} and \ref{t:3} 
is a reduction to the {\SSCDS} problem.
This extends some techniques used previously for $2$-edge connectivity problems 
to $2$-node-connectivity problems. Specifi\-cally, 
Theorem~\ref{t:2} uses ideas from the papers of Basavaraju et al. \cite{BFGM} and Byrka, Grandoni and Ameli \cite{BGA}, while
Theorem~\ref{t:3} uses ideas from the papers of Gupta, Krishnaswamy and Ravi \cite{GKR2} and Belgi and Nutov \cite{BN}.

We now briefly survey some literature on our main problems - {\TCA} and {\TCDS}, 
and their edge-connectivity versions -- {\sc Tree/Cactus Augmentation} and {\sc $2$-Edge-Connec\-ted Dominating Set}.

{\sc Tree Augmentation} and {\TCA} admit several 
simple $2$-approximation algorithms even when the edges have costs \cite{FJa,KT}.
The problem of obtaining ratio below $2$ for the min-size case for these problems was posed by Khuller \cite[p.~263]{K-sur}.
Subsequently, the {\sc Tree Augmentation} problem  was studied in many papers 
\cite{N,CJR,N,EFKN-TALG,CKKK,MN,CN,KN-TAP,BFGM,A,FGKS,N-T,GKZ},
culminating with the current best ratio $1.393$ \cite{CTZ}. 
Very recently, the min-cost version of {\sc Tree Augmentation} was shown to achieve ratio $1.5+\eps$ \cite{TZ2,TZ}. 

In the  {\sc Cactus Augmentation} problem we seek to make a cactus $3$-edge-connected.
{\sc Tree Augmentation} is a particular case, when eve\-ry cycle has length $2$.
Basavaraju et al. \cite{BFGM} gave an approximation ratio preserving reduction from
{\sc Cactus Augmentation} to {\sc SS-CDS} instances with properties 1,2.
Byrka, Grandoni, and Ameli \cite{BGA} showed that such instances admit ratio $1.91$,
thus obtaining the first ratio below $2$ for {\sc Cactus Augmentation}.
The ratio was recently improved to $1.393$ \cite{CTZ}. 

A node subset $S$ is an {\bf $m$-dominating set} if each $v \notin S$ has $m$ neighbors in $S$.  
Problems of finding low size/weight $k$-connected $m$-dominating set were studied both in general graphs and in unit disk graphs. 
In general graphs non-trivial approximation algorithms are known only 
for the case $m \geq k$; c.f. \cite{GK98,F,ZZMD,N-CDS,BN} why the case $m \geq k$ is easier.
For $m<k$, constant ratios are known for unit disk graphs and $k \leq 3$ and $m \leq 2$ \cite{WKAG}, 
and ratio $\tilde{O}(\log^2 n)$ for {\sc $2$-Edge-Connected Dominating Set} in general graphs \cite{BN}.
It was an open problem to obtain a non-trivial ratio for {\sc $2$-Connected Dominating Set} in general graphs.
Note that our result in Corollary~\ref{c:3} is valid for unit node-weights only. 
For general node weights, the problem is still open.

Theorems \ref{t:1}, \ref{t:2}, and \ref{t:3} are proved in Sections \ref{s:1}, \ref{s:2}, and \ref{s:3}, respectively. 

\section{Reducing {\TCA} to {\CFC} (Theorem~\ref{t:1})} \label{s:1} 

\begin{figure}
\centering 
\includegraphics{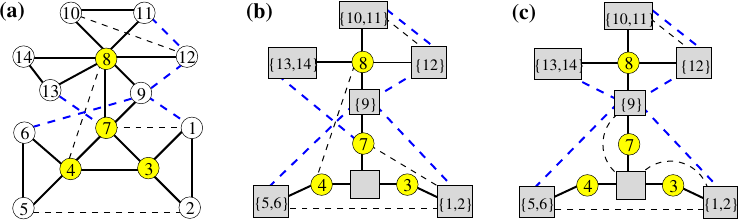}
\caption{Illustration to the reduction of {\TCA} to block-tree augmentation. 
Cut-nodes of $G$ and non-block nodes of $T$ are yellow, 
edges of $G$ and of  $T$ are shown by solid lines, 
and the blue dashed edges show corresponding feasible solutions.
(a,b) The reduction to a block tree. (c) The instance obtained by replacing each link $uv$ by $u'v'$.}
\label{f:bt}
\end{figure}

Given a {\TCA} instance $(G,E)$, the problem reduces by a known reduction (c.f. \cite{K-sur})  
to an appropriate augmentation problem of the block-tree of $G$, see Fig.~\ref{f:bt}(a,b).
A node $x$ is a {\bf cut-node} of a connected graph $G$ if $G \sem \{x\}$ is disconnected. 
A {\bf block} of $G$ is an inclusion maximal subgraph that has no cut-node; note that a bridge is always a block.
The node set of the {\bf block-tree $\TT$ of $G$} is the set of blocks $S$ of $G$ together with the set $X$ of the cut-nodes of $G$, 
and there is an edge between a block in $S$ and a cut-node $x \in X$ if this block contains $x$.
It is known that $\TT$ is indeed a tree and that every leaf of $\TT$ is in $S$. 
Let $\varphi:V(G) \rightarrow V(\TT)$ be a mapping defined by
$\varphi(v)=v$ if $v \in X$ and $\varphi(v)$ is the block that contains $v$ otherwise.
Then $E$ is mapped to the edge set $E'=\{\varphi(u)\varphi(v):uv \in E\}$ on $V(\TT)$,
and {\TCA} is equivalent to the problem of augmenting $\TT$ by 
a min-size edge set $J \subs E'$ such that $\TT \cup J$ has no cut-node in $X$;
see \cite{K-sur} for more details. 

The obtained augmentation problem on the block-tree can be formulated as a problem of covering pairs of sets by edges.
We say that $A \subset V(\TT)$ is a {\bf tight set} if there is $x \in X$ (the cut-node of $A$) such that $A$ is
a non-empty union of some but not all connected components of $\TT \sem \{x\}$; 
note that $A^*=V(\TT) \sem (A \cup \{x\})$ is also a tight set,
and we will call such a pair $(A,A^*)$ a {\bf tight setpair}, with cut node $x$. 
We say that {\bf an edge $e \in E'$ covers $(A,A^*)$} if $e$ has one end in $A$ and the other in $A^*$.
Then $J \subs E'$ is a feasible solution to this block-tree augmentation instance iff $J$ covers all tight setpairs. 
A simple but key obsrvation for our reduction is the following. 

\begin{figure}
\centering 
\includegraphics{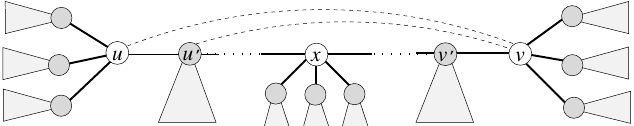}
\caption{Illustration to the proof of Lemma~\ref{l:cov}; block nodes are shown by gray circles.}
\label{f:cov}
\end{figure}

\begin{lemma} \label{l:cov}
Let $uv \in E'$ with $u \in X$, and let $u' \in S$ be the neighbor of $u$ on the $uv$-path in $\TT$.
Then the edge $u'v$ covers the same tight setpairs as $uv$. 
\end{lemma}
\begin{proof}
For illustration see Fig.~\ref{f:cov}.  
Consider a tight setpair $(A,A^*)$ with cut-node $x \in X$. 
One can see that if $uv$ or $u'v$ covers $(A,A^*)$, then
$x$ must be an internal node of the $u'v$-path in $\TT$, and that
$u \in A$ iff $u' \in A$. 
This implies that $uv$ covers $(A,A^*)$ if and only if $u'v$ covers $(A,A^*)$.
\end{proof}

By repeatedly replacing $uv$ by $u'v$ as in Lemma~\ref{l:cov}, 
we obtain an equivalent instance in which $E'$ is an edge set on $S$, see Fig.~\ref{f:bt}(c).
Then $J \subs E'$ covers all tight setpairs iff $J$ covers
the family $\{(A \cap S,A^* \cap S): (A,A^*) \mbox{ is tight}\}$ of the projections of the tight setpairs on $S$;
and since $A^* \cap S=S \sem A$, this is equivalent to $J$ covering the set family $\FF=\{A \cap S: A \mbox{ is tight}\}$.

Summarizing, given a {\TCA} instance $(G,E)$, we construct a {\CFC} instance 
with graph $(S,E')$ and set family $\FF$ as follows.
\begin{enumerate}
\item
Let $(\TT,\varphi)$ be the block-tree of $G$ and let $E' \gets \{\varphi(u)\varphi(v):uv \in E\}$.
\item
Replace every edge $uv \in E'$ by an edge $u'v'$, 
where $u'=u$ if $u \in S$ and $u'$ is the neighbor of $u$ on the $uv$-path in $\TT$ otherwise;
$v'$ is defined similarly. 
After this step, $E'$ is an edge set on $S$. 
\item
The set family we need to cover is $\FF=\{A \cap S: A \mbox{ is tight}\}$.
\end{enumerate}

\begin{figure}
\centering 
\includegraphics{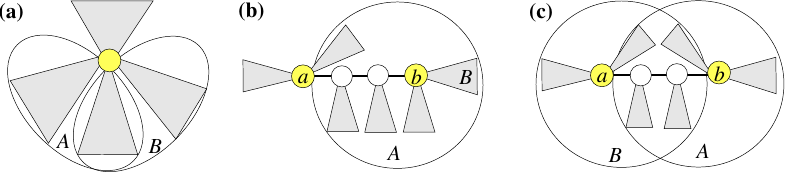}
\caption{Illustration to the proof of Lemma~\ref{l:tight}.}
\label{f:split}
\end{figure}

\begin{lemma} \label{l:tight}
$\FF=\{A \cap S: A \mbox{ is tight}\}$ is a symmetric crossing family. 
\end{lemma}
\begin{proof}
$\FF$ is symmetric since if $A$ is tight then so is $A^*$, and since $A^* \cap S = S \sem (A \cap S)$.
We show that $\FF$ is a crossing family. 
Let $A_S,B_S \in \FF$ cross. 
Let $A,B$ be tight sets with cut-nodes $a,b$ respectively, such that $A \cap S=A_S$ and $B \cap S=B_S$.
Since $A_S,B_S$ cross, so are $A,B$.
 If $a=b$ or if $a \neq b$ and one of $A,B$ contains the other, see Fig.~\ref{f:split}(a,b), then 
$A \cap B, A \cup B$ are both tight and thus $A_S \cap B_S,A_S \cup B_S \in \FF$.
The remaining case when $a \in B$ and $b \in A$ is not possible, since then $A,B$ do not cross, 
see Fig.~\ref{f:split}(c).
\end{proof}

This concludes the proof of Theorem~\ref{t:1}.

\section{Reducing {\CFC} to {\SSCDS} with properties 1,2 (Theorem~\ref{t:2})} \label{s:2} 

Let $\FF$ be a set family and $J$ an edge set on $V$. 
For $A,X,Y \subs V$ we say that $A$ {\bf separates $X$ from $Y$} if $X \subs A$ and $Y \subs V \sem A$.
We use a similar terminology for $X,Y \subs V \cup J$, meaning that $A$ separates 
the nodes and endnodes of the edges in $X$ from
the nodes and endnodes of the edges in $Y$. 
If  there is $A \in \FF$ that separates $X$ from $Y$ or $Y$ from $X$, then $X,Y$ are {\bf $\FF$-separable}, 
and $X,Y$ are {\bf $\FF$-inseparable} otherwise.
For $s,t \in V$ let $\FF^{st}=\{A \in \FF:s \in A,t \in V \sem A\}$. 
The relation ''$s,t \in V$ are $\FF$-inseparable'' is an equivalence, 
and its equivalence classes are called {\bf $\FF$-atoms}.
W.l.o.g. we may contract each $\FF$-atom into a single node, and assume that all $\FF$-atoms are singletons. 
Note that then any $s,t \in V$ are $\FF$-separable. 
The {\bf separability graph} $H=(U,I)$ of $\FF,J$ has node set $U=V \cup J$ 
and edge set $I=\{fz: f \in J \mbox{ and } z \in U \mbox{ are inseparable}\}$,
and we denote by $H^{st}=H[\{s,t\} \cup J]$ the graph obtained by removing $V \sem \{s,t\}$ from $H$. 

\begin{lemma} \label{l:ring}
Let $\FF$ be a symmetric crossing family with all $\FF$-atoms being singletons,
let $J$ be an edge set $J$ on $V$, and let $s,t \in V$. 
Let $H$ be the separability graph of $\FF,J$. 
Then $J$ covers $\FF^{st}$ iff $H^{st}$ has an $st$-path. 
(Namely, $J$ does not cover some $A \in \FF^{st}$ iff $H^{st}$ has no $st$-path.)
\end{lemma}
\begin{proof}
Suppose that $J$ does not cover some $A \in \FF^{st}$. 
Let $J_s$ be the set of edges in $J$ contained in $A$.
No edge in $J$ covers $A$, so $H^{st}$ has no edge between 
$J_s \cup \{s\}$ and $J \sem J_s \cup \{t\}$. This contradicts that $H^{st}$ has an $st$-path. 

Suppose that $H^{st}$ has no $st$-path.
Let $J_s$ be the set of nodes of the connected component of $H^{st}$ that contains $s$, see Fig.~\ref{f:ab-kc}(a).
We now prove that for any $g \in (J \sem J_s) \cup \{t\}$,
$\FF$ contains a set that separates $J_s$ from~$g$. 
Let $f_0,f_1,\ldots$ be an ordering of $J_s$, where $f_0=s$, 
such that each $f_i$ is adjacent in $H$ to some $f_j$ with $j <i$;
since $H[J_s]$ is connected, such an ordering exists. 
For any $f_i$ there is $A_i \in \FF$ that separates $f_i$ from $g$.
Note that if $A$ separates $X$ from $Z$ and $B$ separates $Y$ from $Z$, then 
$A \cup B$ separates $X \cup Y$ from $Z$, 
and if $A \cap B=\empt$ then $A$ separates $X$ from $Y$;
see Fig.~\ref{f:ab-kc}(b,c).
Let $Z=\{g\}$.
Apply the above on $A=A_0,B=A_1$ and $X=\{f_0\}=\{s\},Y=\{f_1\}$.  
Since $f_0$ and $f_1$ are inseparable, $A \cap B \neq \empt$. 
Thus $A,B$ cross,
hence the set $A \cup B=A_0 \cup A_1$ belongs to $\FF$ and separates $\{f_0,f_1\}=\{s,f_1\}$ from $g$.
By an identical argument applied on $A=A_0 \cup A_1,B=A_2$ and
$X=\{f_0,f_1\},Y=\{f_2\}$ we get that $(A_0 \cup A_1) \cup A_2 \in \FF$ separates $\{f_0,f_1,f_2\}$ from $g$.
By induction we get that $\cup_{i}A_i \in \FF$ separates $J_s$ from $g$.

\begin{figure}
\centering 
\includegraphics{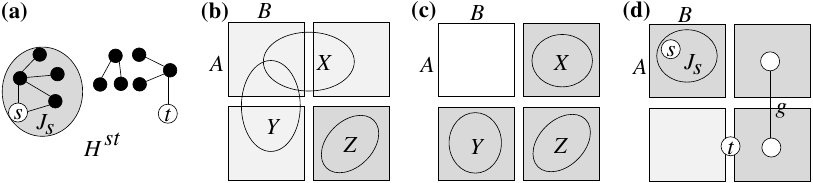}
\caption{Illustration to the proof of Lemma~\ref{l:ring}. 
Dark gray parts are non-empty, 
light gray parts may or may not be empty.
In (d) we may have $t \in V \sem (A \cup B)$ or $t \in B \sem A$.}
\label{f:ab-kc}
\end{figure}

Let now $A$ be an inclusion minimal set in $\FF$ that separates $J_s$ from $t$,
and note that $A \in \FF^{st}$.
We claim that $J$ does not cover $A$.
Suppose to the contrary that some $g \in J$ covers $A$. 
Since $g \in J \sem J_s$, there is $B \in \FF$ that separates $J_s$ from $g$, see Fig.~\ref{f:ab-kc}(d).
As $g$ has both ends in $V \sem B$ and covers $A$, 
it has one end in $V \sem (A \cup B)$ and the other in $A \sem B$.
Thus $A,B$ cross, hence $A \cap B \in \FF$. 
Then the set $A \cap B \in \FF^{st}$ and separates $J_s$ from $t$. 
But $A \cap B \subsetneq A$, contradicting the minimality of $A$.  
\end{proof}

If $\FF$ is symmetric and crossing, then every minimal set in $\FF$ is an $\FF$-atom,
and w.l.o.g. is a singleton; let $R_\FF$ denote the set of these singletons. 
Note that $J$ covers $\FF$ iff $J$ covers $\FF^{st}$ for all $s,t \in R_\FF$.
Thus Lemma~\ref{l:ring} implies:

\begin{corollary} \label{c:cr}
Let $\FF$ be a symmetric crossing family with all $\FF$-atoms being singletons, 
and let $J$ be an edge set on $V$.
Then $J$ covers $\FF$ iff $J$ is a connected $R_\FF$-dominating set in the separability graph $H$ of $\FF,J$. 
\end{corollary}

Corollary~\ref{c:cr} implies Theorem~\ref{t:2}.
Given a {\CFC} instance $(G=(V,E),\FF)$ 
construct a {\SSCDS} instance $(H,R=R_\FF)$ by removing from the separability graph of $\FF,E$ the nodes in $V \sem R_\FF$. 
By Corollary~\ref{c:cr}, $J \subs E$ is a feasible solution to the {\CFC} instance 
iff $J$ is a feasible solution to the constructed {\SS-CDS} instance. 
Note that $H$ satisfies properties $1,2$ from the Introduction:
\begin{enumerate} 
\item 
The neighbors of every terminal $r \in R$ induce a clique, 
since any two edges incident to the same node are inseparable.
\item
Every non-terminal $u \in U \sem R$ has at most $2$ neighbors in $R$, since every edge connects at most $2$ nodes in $R$. 
\end{enumerate}

The {\SSCDS} instance $(H,R)$ can be constructed using $O(|E|^2)$ calls to the oracle, that given $(s,t) \in V \times V$ and $J \subs E$,
returns an inclusion minimal set $C^{st}  \in \FF^{st}$ that is not covered by $J$, or determines that such does not exist. 
We do this for every pair $(s,t)$ with $J=\empt$. Then $s,t$ belong to the same $\FF$-atom iff $C^{st}$ does not exist,
and the minimal sets in $\FF$ are the minimal ones among at most $|V|(|V|-1)$ sets computed. 
Also, given a pair $f=uv$ and $f' =u'v'$, we can check in polynomial time whether $f,f'$ are separable,
by checking whether there is $A \in \FF$ that is not covered by 
$\{f,f'\}$ such that $|A \cap \{s,t\}|=1$ for some $(s,t) \in \{u,v\} \times \{u'v'\}$.


This concludes the proof of Theorem~\ref{t:2}. 

\medskip

We illustrate Lemma~\ref{l:ring} and Corollary~\ref{c:cr} on the problem of augmenting 
a tree $T=(V,E_T)$ by a min-size edge set $J \subs E$ such that $T \cup J$ is $2$-connected. 
Applying the Section~\ref{s:1} reduction, we get that in the {\CFC} instance, $f,g \in J$ are inseparable iff in the block-tree $\TT$ of $T$,
the paths $\TT_f$ and $\TT_g$ between the ends of $f$ and $g$, respectively, share a block-node. 
Since every block-node of $\TT$ corresponds to an edge of $T$, 
this is equivalent to the property that the paths $T_f$ and $T_g$ (of $T$) share an edge. 
Thus we get the following statement, that in the preliminary version \cite{N-W} of this paper was proved directly.

\begin{corollary} \label{c:T}
Let $T=(V,E_T)$ be a tree with leaf set $R$ and $J$ an edge set.
Let $H=(U,I)$ have node set $U=E_T \cup R$ and edge set $I=I_{con} \cup I_{inc}$ where
\begin{eqnarray*}
I_{con} &=& \{fg:f,g \in J, T_f \mbox{ and } T_g \mbox{ share an edge}\} \\ 
I_{inc}  &=& \{rf:f \in J,r \in R, f \mbox {is incident to } r\}
\end{eqnarray*}
Then $T \cup J$ is $2$-connected iff $J$ is a connected $R$-dominating set in $H$. 
\end{corollary}

\section{Reducing {\TCDS} to {\SSCDS} (Theorem~\ref{t:3})} \label{s:3} 

Suppose that we want to solve a generic {\sc $2$-Connected Network Design} problem, defined as follows: \\ 
{\em Given a graph $G=(V,E_G)$ with edge costs $\{c_e:e \in E_G\}$, find a cheapest $2$-connected subgraph 
$(S,F)$ of $G$ that satisfies a prescribed property $\Pi$.} \\
The property $\Pi$ can be viewed as a family of subgraphs of $G$,
but it does not need to be given explicitly.
For example, in the {\TCDS} problem, 
the family $\Pi$ consists of all subgraphs of $G$ whose node set dominates $V$. 
We will assume that {\bf $\Pi$ is monotone}, meaning that if $G' \in \Pi$, then any supergraph of $G'$ is also in $\Pi$. 
We will show a method to reduce this problem to a problem of finding a min-weight {\em connected} subgraph
that satisfies a certain ``relevant'' property in a node weighted graph.
The main advantage is that we transform a $2$-connectivity problem into a plain connectivity problem.
But there are also three disadvantages. 
\begin{enumerate}
\item[(i)]
We get a node weighted problem, and these are usually harder than problems with edge costs. 
\item[(ii)]
The reduction invokes a factor of 
$\sigma=O(\log n \cdot\log\log n\cdot(\log\log\log n)^{3})$ in the approximation ratio, 
due to using the probabilistic spanning tree embedding of \cite{ABN}. 
\item[(iii)]
The approximation ratio is ''in expectation'', without a precise probability guarantee. 
\end{enumerate}

The algorithm of \cite[Theorem~1]{ABN} computes a distribution $\{p_T:T \in {\cal T}\}$ 
of spanning trees of  a graph $G$ with edge costs, with average stretch $\sigma$. 
We draw one tree from this distribution, and it has expected stretch $\si$.
So, we will assume that we have a single spanning tree $T$ with (expected) stretch $\si$, 
namely, that $c(T_f) \leq \si c(f)$ for all $f \in E_G$, where $T_f$ denote the unique path in $T$ between the endnodes of $f$.
From this point, the reduction has two stages: a reduction to a certain ``tree augmentation'' problem, 
that is further reduced to a variant of {\SSCDS} using methods from the previous section. 

Let $T=(V,E_T)$ be a tree and $E$ an edge set on $V$. 
For $J \subs E$ let $T_J=\cup_{f \in J} T_f$ denote the forest formed by the union of the paths $T_f$ of the edges in $J$.
The first reduction is to the following problem.

\begin{center} \fbox{\begin{minipage}{0.98\textwidth}
\underline{{\sc Subtree Augmentation Network Design}} \\
{\em Input:} \ A tree $T=(V,E_T)$, an edge set $E$ on $V$ with costs, and a property~$\Pi$.  \\
{\em Output:} A min-cost edge set $J \subs E$ such that $T_J$ is a tree and 
the graph $T_J \cup J$ is $2$-connected and satisfies $\Pi$.
\end{minipage}} \end{center}

Given an instance $\II_G=(G=(V,E_G),c,\Pi)$ of {\sc $2$-Connected Network Design} 
and a spanning tree $T=(V,E_T)$ in $G$ with stretch $\si$,
we compute a solution $J$ to the {\sc Subtree Augmentation Network Design} instance 
$\II_T=(T,E=E_G \sem E_T,\Pi)$, and return $T_J \cup J$. 
Clearly, if $J$ is a feasible solution to $\II_T$ then $T_J \cup J$ is a feasible solution to $\II_G$.
The next lemma shows the other direction, and also establishes the fee of the reduction.

\begin{lemma} \label{l:si}
If $(S,F)$ is a feasible solution to $\II_G$ then $J=F \sem E_T$ is a feasible solution to $\II_T$.
Furthermore, $c(T_J \cup J) \leq (\si+1)c(F)$.
\end{lemma}
\begin{proof}
For the first statement we will show that: (i) $T_J$ is a tree; (ii) $T_J \cup J$ is $2$-connected; (iii) $(S,F)$ is a subgraph of $T_J \cup J$.
Note that (iii) implies that $T_J \cup J$ satisfies $\Pi$, since we assume that the property $\Pi$ is monotone. 

We prove that $T_J$ is a tree.
Suppose to the contrary that $T_J$ has distinct connected components $C,C'$. 
Let $e$ be the first edge on the path in $T$ from $C$ to $C'$.
No edge in $J$ has exactly one end in $C$, hence $(T \cup F) \sem \{e\}$
is disconnected, and one of its connected components contains $C$ and the other contains $C'$. 
This implies that $(S,F \sem \{e\})$ is disconnected, contradicting that $(S,F)$ is $2$-connected.

To see that $T_J \cup J$ is $2$-connected and that $(S,F)$ is a subgraph of $T_J \cup J$, 
note that the graph $T_J \cup J$ is obtained from the $2$-connected graph $(S,F)$
by sequentially adding for each $f \in J$ the path $T_f$.
Adding a simple path $P$ between two distinct nodes of a $2$-connected graph  
results in a $2$-connected graph; this is so also if $P$ contains some edges of the graph. 
It follows by induction that  $(S,F) \cup ( \cup_{f \in J} T_f)$ is $2$-connected.

To see that $c(T_J \cup J) \leq (\si+1)c(F)$, note that $J \subs F$ implies $c(J) \leq c(F)$, and that 
$c(T_J) \leq \sum_{f \in J} c(T_f) \leq  \sum_{f \in J} \si c(f) = \si c(J)$.
\end{proof}

Now consider the {\TCDS} problem, that seeks
a $2$-connected subgraph $(S,F)$ of $G$ that dominates $V$ and has minimum number $|F|$ of {\em edges};
as was mentioned in the Introduction, this is equivalent to minimizing the number of nodes, up to a factor of $2$.
We reduce this problem to {\sc Subtree Augmentation Network Design} with the property $\Pi_{dom}$ that 
$T_J$ should dominate $V$. This invokes a factor of of $\si+1$ in the approximation ratio, by Lemma~\ref{l:si}.
Now we reduce the later problem to {\sc SS-CDS}. 
For this reduction we need the following immediate consequence from Corollary~\ref{c:T}.

\begin{corollary} \label{c:red}
Let $T=(V,E_T)$ be a tree and $J$ an edge set. 
Let $H_{con}$ be a graph with node set $J$ and edge set 
$I_{con}=\{fg:T_f \mbox{ and } T_g \mbox{ share an edge}\}$.
Then $T_J \cup J$ is $2$-connected iff $H_{con}$ is connected.
\end{corollary}

\begin{figure}
\centering 
\includegraphics{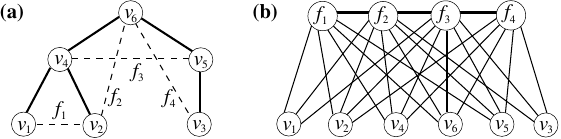}
\caption{Illustration to the construction of a {\sc SS-CDS} instance. 
(a) Edges in $T$ are shown by bold lines and edges in $E$ by dashed lines.
(b) Edges of $I_{con}$ are shown by bold lines and edges in $I_{dom}$ are shown by thin lines.}
\label{f:dom}
\end{figure}

Given a  {\sc Subtree Augmentation Network Design} instance $\II=(T=(V,E_T),E,\Pi_{dom})$,
construct a {\sc SS-CDS} instance $\II'=(H=(U,I),R)$ as follows, see Fig.~\ref{f:dom}.
Let $D_f$ denote the nodes dominated by $T_f$ in $G=T \cup E$;
e.g., in Fig.~\ref{f:dom}, where $V=\{v_1, \ldots,v_6\}$, we have:
$D_{f_1}=V \sem \{v_3\}$, 
$D_{f_2}=D_{f_3}=V$, and 
$D_{f_4}=V \sem \{v_1\}$.
The graph $H$ has node set $U=V \cup E$ and edge set $I=I_{con} \cup I_{dom}$ 
where $I_{con}$ and $I_{dom}$ are defined by: 
\begin{eqnarray*}
I_{con}  &=& \{fg:f,g \in E, T_f \mbox{ and } T_g \mbox{ share an edge}\} \\
I_{dom} &=& \{fv: f \in E, v \in D_f\}
\end{eqnarray*}
The set of terminals of $H$ is $R=V$.

Let $J \subs E$. Then $J \cup T_J$ is $2$-connected if and only if $H[J]$ is connected;
this follows from Corollary~\ref{c:red}.
By the construction, $T_J$ dominates $V$ in the graph $G=T \cup E$ if and only if $J$ dominates $V$ in $H$. 
Thus $F$ is a feasible solution to the {\sc Subtree Augmentation Network Design} instance $\II$ iff
$J$ is a feasible solution to the constructed {\sc SS-CDS} instance $\II'$.

To show overall (expected) approximation ratio $O(\si \log^3 n)$ we show that {\sc SS-CDS} admits ratio $O(\log^3 n)$.
We give a reduction to the {\sc Group Steiner Tree} ({\sc GST}) problem: 
given a graph with edge costs and a collection of subsets (groups)
of the node set, find a min-cost subtree that contains a node from every group. 
Given a {\SSCDS} instance $(H,R)$ obtain an equivalent {\sc GST} instance {\em with unit node weights} as follows:
for every $r \in R$, introduce a group $S_r$ that consists of the neighbors of $r$ in $H$, and then remove $r$. 
Since feasible solutions are trees, we may minimize the number of edges, and get a {\sc GST} instance with {\em unit edge costs}.
By \cite{GKR} {\sc GST} (with edge costs) admits ratio $O(\log k \log M \log N)=O(\log^3 n)$, 
where $k$ is the number of groups, $M$ is the maximum group size, and $N$ is the number of nodes;
in our case, $k=|R|=|V|$, $M=\max_{r \in R} \deg_H(r) \leq |E|$, and $N=|E|$.

This concludes the proof of Theorem~\ref{t:3}. 

\section{Additional applications} \label{s:appl}

Here we will apply our techniques on two problems considered in \cite{CK}.
Our approximation ratios are in expectation and also slightly worse than those in \cite{CK}, 
but the algorithms and the proofs 
are very simple (except that we use a highly non-trivial tree embedding of \cite{ABN}). 
The first problem we consider is: 

\begin{center} \fbox{\begin{minipage}{0.98\textwidth}
\underline{{\sc $2$-Connected $k$-Subgraph}} \\
{\em Input:} \ \ A graph $G=(V,E)$ with edge costs $\{c_e:e \in E\}$ and an integer $k$. \\
{\em Output:} A min-cost $2$-connected subgraph $(S,F)$ of $G$ with $|S| \geq k$. 
\end{minipage}} \end{center}

We give a simple proof that this problem admits (expected) approximation ratio $O(\si \log k)$.
Consider the {\sc Subtree Augmentation Network Design} instance $(T=(V,E_T),E,c,\Pi_k)$, 
obtained by applying the reduction from Section~\ref{s:3}, that incurs a factor of $\si+1$ in the ratio;
here $\Pi_k$ is the property that $T_J$ should have at least $k$ nodes. 
Namely, we need to find a min-cost edge set $J \subs E$ such that $T_J \cup J$ is $2$-connected and has at least $k$ nodes.
Note that $T_J \cup J$ has at least $k$ nodes if and only if $T_J$ has at least $k-1$ edges.
Next, construct a node weighted graph $H$ with node set $E \cup E_T$ and edge set 
$I_{con} \cup I_{inc}$, where 
\begin{eqnarray*}
I_{con} &=& \{fg:f,g \in E, T_f \mbox{ and } T_g \mbox{ share an edge}\} \\ 
I_{inc} &=& \{fg: f \in E,g \in E_T, f  \in T_g\}
\end{eqnarray*}
The set of terminals of $H$ is $R=E_T$ and they have zero weights, 
and non-terminals have weights $w(f)=c(f)$ for all $f \in E$.
In $H$, we need to find a min-weight node subset $J \subs E$ such that $H[J]$ is connected 
and $J$ has at least $k-1$ neighbors in $R$.
This is an instance of the weighted {\sc Partial {\SSCDS}} problem, that by \cite{BN} is harder than {\sc GST}. 
However, note that the obtained instance has Property~1 - the neighbors of every terminal $r \in R$ induce a clique.
Thus the problem is equivalent to finding a min-weight subtree with at least $k-1$ terminals, and
it admits ratio $O(\log k)$, by \cite{BHL,KSS}. The overall (expected) ratio is $O(\si \log k)$.

Now we give a simple proof that the following more general problem admits (expected) approximation ratio $O(\si \log n)$.

\begin{center} \fbox{\begin{minipage}{0.98\textwidth}
\underline{{\sc $2$-Connected Quota Subgraph}} \\
{\em Input:} \ A graph $G=(V,E)$ with edge costs $\{c_e:e \in E\}$, node profits $\{p_v:v \in V\}$, and a quota $Q$. \\
{\em Output:} A min-cost $2$-connected subgraph of $G$ that has profit at least $Q$.
\end{minipage}} \end{center}

We may consider the ``rooted'' version of the problem,
when the subgraph should contain a specified node $s$ of zero profit, c.f. \cite{CK}.
We apply the reduction from the previous section with stretch $\si$ on the costs, and root $T$ at $s$.
For every node $v \neq s$, its parent edge in $T$ gets the profit $p(v)$ of $v$. 
Next, consider the corresponding node weighted problem defined by the same graph $H$ as in the 
{\sc $2$-Connected $k$-Subgraph} problem, with weights $w(f)=c(f)$ and profits $p(f)=0$ for all $f \in E$,
while every node in $H$ that corresponds to an edge $e$ of $T$ has cost zero and inherits the profit of $e$.
Here we need to find a minimum edge weight subtree of node profit at least $Q$, and
it admits ratio $O(\log n)$, by \cite{BHL,KSS}. The overall ratio is $O(\si \log n)$.

\medskip

A similar method can be used to obtain a bi-criteria approximation 
for the {\sc $2$-Connected Budgeted Subgraph} problem, where the goal is to maximize node profit 
under edge cost budget constraints.
The reduction is the same as above, but we need to find a maximum profit subtree of weight at most a given budget $B$.
This problem admits a bi-criteria approximation $\left(1+\eps,\Omega\left(\f{\eps^2}{\log n}\right)\right)$ \cite{BHL}, 
giving the overall approximation $\left((1+\eps)\si,\Omega\left(\f{\eps^2}{\log n}\right)\right)$.
But for this problem the algorithm of \cite{CK},
that achieves a bi-criteria approxi\-mation $\left(2+\eps,\Omega\left(\f{\eps}{\log^2n}\right)\right)$ for any $\eps \in (0,1]$,
has a smaller budget violation.

\medskip \medskip

\noindent
{\bf Acknowledgment.} I thank anonymous referees for many useful comments.


\end{document}